\documentclass[12pt]{article}
\usepackage{amsfonts,amsmath,amssymb,epsf}
\usepackage{graphicx,color}
\usepackage{wrapfig}
\newcommand{\al}{\alpha'}

\newcommand{\Slash}[1]{{\ooalign{\hfil$#1$\hfil\crcr\raise.167ex\hbox{/}}}}
\newcommand{\be}{\begin{equation}}
\newcommand{\ba}{\begin{eqnarray}}
\newcommand{\ea}{\end{eqnarray}}
\newcommand{\ee}{\end{equation}}

\newcommand{\we}{\wedge}

\newcommand{\f}{\frac}

\newcommand{\vp}{\varphi}

\newcommand{\ti}{\tilde}

\newcommand{\no}{\nonumber \\}

\newcommand{\ep}{\epsilon}

\def\al{\alpha'}

\def\ep{{\epsilon}}

\def\f#1#2{{\frac{#1}{#2}}}

\def\f {\frac}
\def\ti{\tilde}

\def\we{\wedge}

\begin{document}

\begin{flushright}
TAUP-2950/12

KEK-CP-274

NSF-KITP-12-064
\end{flushright}

\vspace{0.1cm}

\begin{center}

{\LARGE
A new large-$N$ limit 
and 

the planar equivalence outside the planar limit

  }
\end{center}
\vspace{0.1cm}
\vspace{0.1cm}
\begin{center}

         Mitsutoshi F{\sc ujita}$^{a}$,  
         Masanori H{\sc anada}$^{bc}$  
and 
 Carlos H{\sc oyos}$^{d}$ 

${}^a$ {\it Department of Physics, University of Washington, 
 Seattle, WA 98195-1560, USA}\\

${}^{b}$ {\it KEK Theory Center, High Energy Accelerator Research Organization (KEK), 

		Tsukuba 305-0801, Japan}\\

$^c$ {\it Kavli Institute for Theoretical Physics, University of California,

Santa Barbara, CA 93106-4030, USA}\\

 ${}^{d}$
{\it Raymond and Beverly Sackler School of Physics and Astronomy,  Tel-Aviv University, 

Ramat-Aviv 69978, Israel}\\

\end{center}

\begin{center}
  {\bf Abstract}
\end{center}
We consider a new large-$N$ limit, in which the 't Hooft coupling grows with $N$. 
We argue that a class of large-$N$ equivalences, which is known to hold in the 't Hooft limit, 
can be extended to this very strongly coupled limit.  
Hence this limit may lead to a new way of studying corrections to the 't Hooft limit, while keeping nice properties of the latter.   
As a concrete example, 
we describe large-$N$ equivalences between the ABJM theory and its orientifold projection. 
The equivalence implies that operators neutral under the projection symmetry have the same correlation functions in two theories at large-$N$. 
Usual field theory arguments are valid when 't Hooft coupling $\lambda\sim N/k$ is fixed and observables can be computed by using a planar diagrammatic expansion. With the help of the AdS/CFT correspondence, we argue that the equivalence extends to stronger coupling regions, $N\gg k$, including the M-theory region $N\gg k^5$. 
We further argue that the orbifold/orientifold equivalences between certain Yang-Mills theories can also be generalized. 
Such equivalences can be tested both analytically and numerically.  
Based on calculations of the free energy, we conjecture that the equivalences hold because planar dominance persists beyond the 't Hooft limit. 

\newpage

\section{Introduction and summary}

The 't Hooft large-$N$ limit (planar limit) of gauge theories \cite{'tHooft:1973jz}, in which the 't Hooft coupling is fixed,  
plays a prominent role in theoretical particle physics. The diagrammatic $1/N$ expansion of gauge theories in the 't Hooft limit can be regarded as the genus expansion 
of a string theory.  It is then expected that the large-$N$ limit of a gauge theory provides a nonperturbative formulation of a string theory. 
Indeed the AdS/CFT duality \cite{Maldacena:1997re} (or more generally the gauge/gravity duality \cite{Itzhaki:1998dd}) provide us with concrete realizations. 

In the strict large-$N$ limit, only planar diagrams survive. In the gravity language, this corresponds to the classical string limit. 
Theories drastically simplify in this limit, and surprising properties hold even in theories without a gravity dual. 
In particular, seemingly very different theories become equivalent. The first examples are the Eguchi-Kawai equivalence, which claims that
certain gauge theories and matrix models become equivalent \cite{Eguchi:1982nm}, 
and the equivalence between pure Yang-Mills theories with $U(N)$, $O(2N)$ and $USp(2N)$ gauge groups \cite{Lovelace:1982hz}. 
Today these equivalences are understood as special cases of the {\it orbifold equivalence} and {\it orientifold equivalence} \cite{Kachru:1998ys}, 
which were found soon after the discovery of the AdS/CFT correspondence. 
The equivalences imply that, when one considers two theories related by the orbifold or orientifold projection, 
operators neutral under the projection symmetry have the same correlation functions in the two theories. 
There are equivalences for theories with and without gravity duals \cite{Bershadsky:1998cb,Kovtun:2004bz}, 
and they also have valuable applications in non-supersymmetric theories \cite{Schmaltz:1998bg,Strassler:2001fs,Armoni:2003gp}, 
including realistic large-$N$ QCD at finite density \cite{Cherman:2010jj,Hanada:2011ju,
Cherman:2011mh,Hidaka:2011jj}\footnote{
See also 
\cite{Cohen:2003kd,Toublan:2005zu} for earlier related works.  
Although a proof given in \cite{Cherman:2010jj,Hanada:2011ju} is applicable only to all orders in perturbation theory, 
there is fairly good evidence that the equivalence holds nonperturbatively; see 
\cite{Hanada:2011ju,Cherman:2011mh,Hanada:2012nj}. For more recent work on this topic see \cite{Blake:2012dp}. 
} and confinement in pure Yang-Mills theory \cite{Unsal:2007vu,Shifman:2008ja,Unsal:2008ch}. 
Therefore it is important to understand these equivalences further. 
In particular, it is interesting to see if equivalences can be generalized outside the 't Hooft limit. 
In this paper, we argue that such equivalences indeed can be valid even outside the 't Hooft limit, in a regime where the 't Hooft coupling grows with $N$, at least for a class of theories. 

The key observation comes from the ${\cal N}=6$ supersymmetric $U(N)\times U(N)$ Chern-Simons-matter theory with level $k$, 
which has been proposed by Aharony, Bergman, Jafferis and Maldacena (ABJM) as the theory of $N$ M2 branes on a $\mathbb{Z}_k$ orbifold \cite{Aharony:2008ug}. 
When $N\gg k^5$ the ABJM theory was conjectured to have a holographic dual description in terms of M-theory on $AdS_4\times S^7/{\mathbb Z}_k.$ Since the 't Hooft coupling of the theory is $\lambda=N/k$, this is not the 't Hooft limit\footnote{Fixed-$k$ large-$N$ limit has also been studied in other theories. See~\cite{Kiritsis:2010xc}.}. 
On the other hand, at $k\ll N\ll k^5$, which includes the 't Hooft limit (with an $O(N^0)$ but strong 't Hooft coupling constant) 
the theory is dual to the reduction of M-theory on the modded circle, type IIA superstring theory on $AdS_4\times {\mathbb C}P^3$. 
According to the ABJM proposal, the large-$N$ behavior of the ABJM theory captures the tree level properties of gravity, both in the type IIA and M-theory regions. 
It enables us to make nontrivial statements on the ABJM theory via the AdS/CFT correspondence, by studying the gravity side. 
In \cite{Hanada:2011yz}, it has been pointed out that there are orbifold equivalences that can be seen on the gravity side \cite{Kachru:1998ys} and that extend to the M-theory region 
without any modification, and that a new orbifold equivalence exists in the M-theory region, which relates $U(kN)_1\times U(kN)_{-1}$ and $U(N)_k\times U(N)_{-k}$ theories. Furthermore in \cite{Hanada:2011zx} it has been shown that this equivalence can naturally be derived on the field theory side if we assume 
mirror symmetry, the equivalence can then be understood as the usual type of orbifold equivalence but between the mirror theories. 

\begin{figure}[htbp]
   \begin{center}
   \scalebox{0.6}{
     \includegraphics[height=6cm]{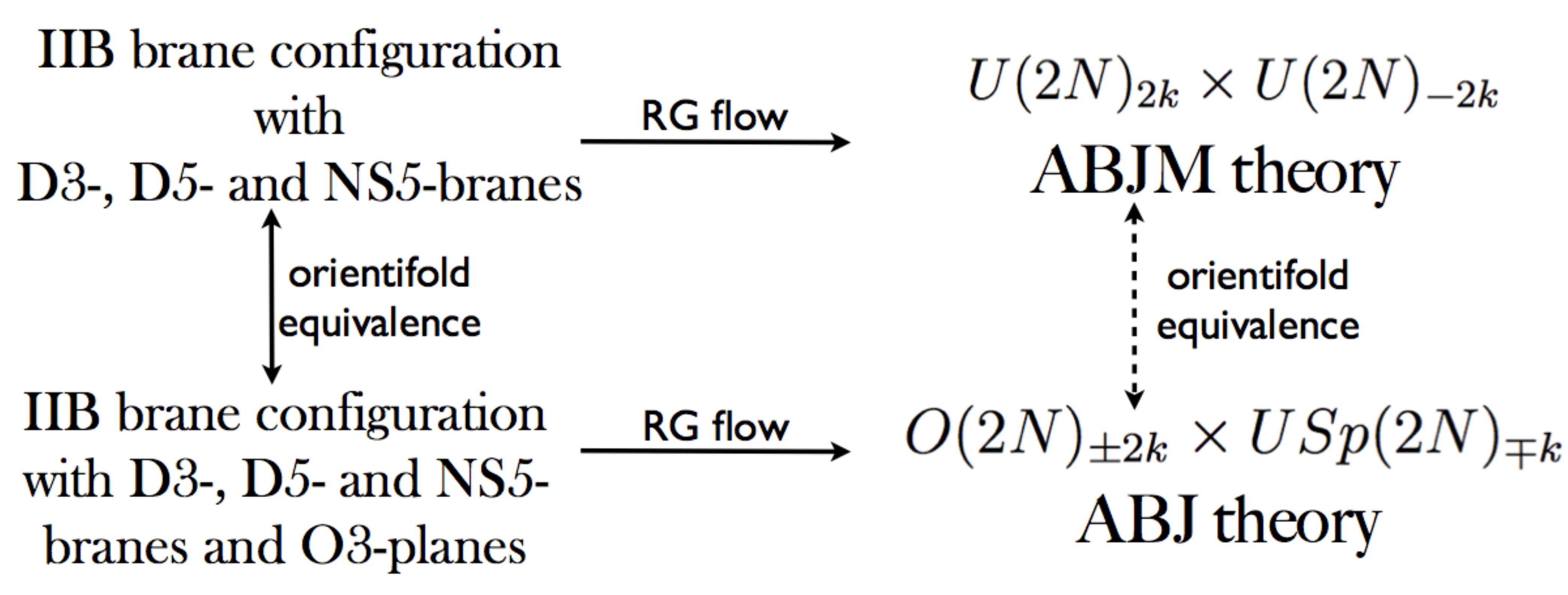}}
   \end{center}
   \caption{UV theories of the ABJM and ABJ theories are related by the orientifold projection and are equivalent.  
   They flow to IR fixed points, ABJM and ABJ theories, which are again related by the orientifold projection. 
   This suggests that the equivalence between ABJM and ABJ, although this argument has a subtlety explained in section~\ref{sec:brane_construction}. 
  }
\label{fig:equivalence_RG}
\end{figure}

\begin{figure}[htbp]
   \begin{center}
   \scalebox{0.6}{
     \includegraphics[height=6cm]{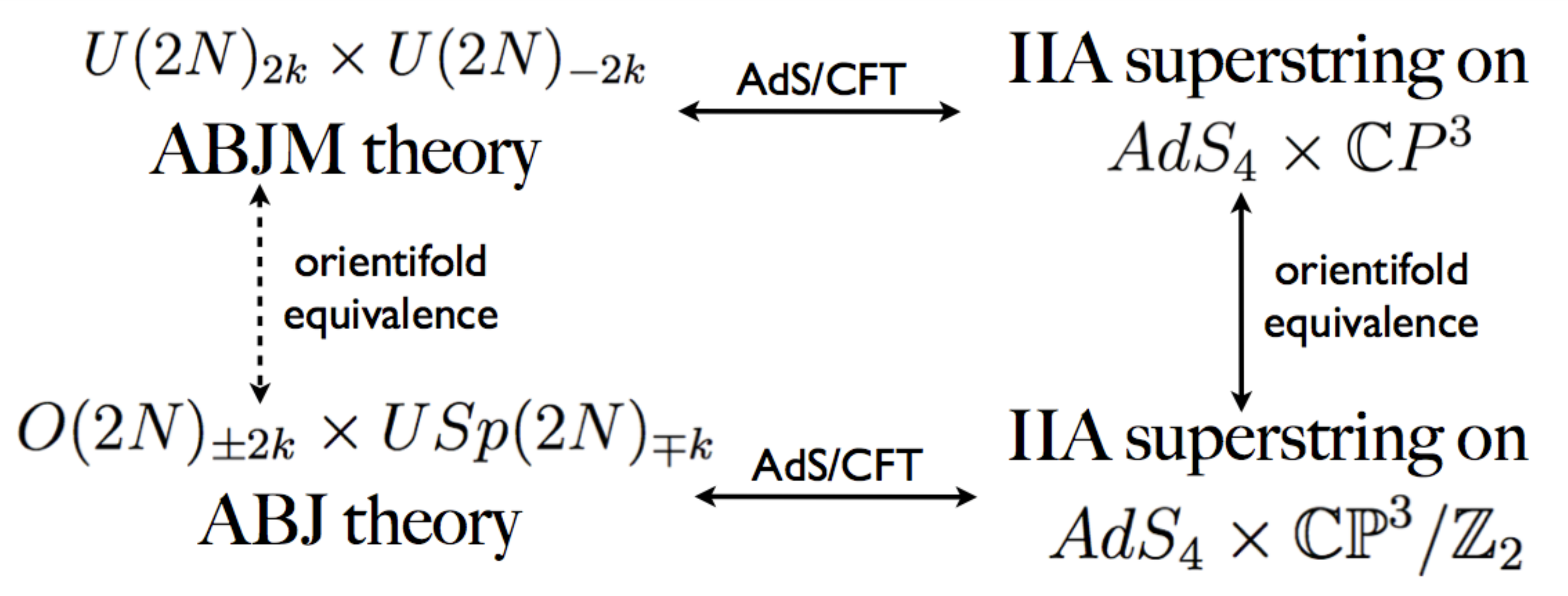}}
   \end{center}
   \caption{The orientifold equivalence in the IIA superstring region ($k\ll N\ll k^5$). The equivalence in the gravity side can be translated into the gauge theory side via the AdS/CFT duality. In the planar limit ($\lambda=N/k$ fixed) the equivalence can also be shown directly in the gauge theory side, without referring to the gravity side. 
   See section~\ref{sec:gravity} for details. 
  }
\label{fig:equivalence_IIA}
\end{figure}

\begin{figure}[htbp]
   \begin{center}
   \scalebox{0.6}{
     \includegraphics[height=6cm]{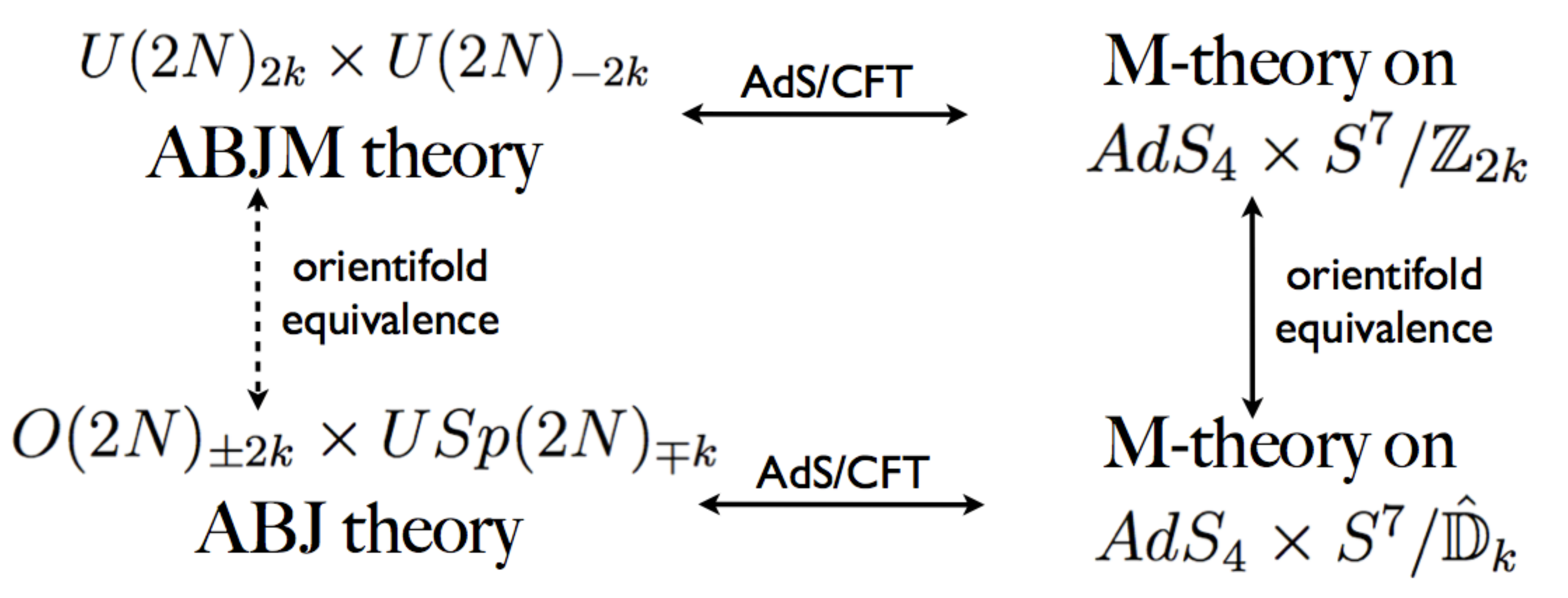}}
   \end{center}
   \caption{The orientifold equivalence in the M-theory region ($k^5\ll N$, $N\to\infty$). The equivalence in the gravity side can be translated into the gauge theory side via the AdS/CFT duality. Usual proof for the 't Hooft limit does not apply in the gauge theory side. See section~\ref{sec:gravity} for details. }
\label{fig:equivalence_M}
\end{figure}

In this paper we consider yet another equivalence, which is probably more familiar to many of the readers:  
the equivalence between the $U(2N)_{2k}\times U(2N)_{-2k}$ ABJM theory 
and its orientifold projection, $O(2N)_{\pm 2k}\times USp(2N)_{\mp k}$ (ABJ model) \cite{Aharony:2008gk,Hosomichi:2008jb},  
\begin{equation}
U(2N)_{2k}\times U(2N)_{-2k}\to O(2N)_{\pm 2k}\times USp(2N)_{\mp k}.\label{EQU9}  
\end{equation}
As we will see in section~\ref{sec:brane_construction}, these theories are the low-energy fixed points 
of type IIB brane configurations which are equivalent through the orientifold equivalence (Fig.~\ref{fig:equivalence_RG}). 
In the 't Hooft limit of the ABJM theory, where $\lambda=N/k$ is fixed, the equivalence in the four-dimensional theory (UV) guarantees the equivalence at the fixed points (IR), namely between the
ABJM and ABJ theories. 
The equivalence in the 't Hooft limit immediately follows from previously known field theory techniques,  
and can also be shown by using the IIA superstring description, thanks to the AdS/CFT duality (Fig.~\ref{fig:equivalence_IIA}). 
On the other hand, when $k$ is smaller than $O(N^1)$, the fixed point is outside the planar region of the UV theory,  
and so it is not guaranteed a priori that the two four-dimensional theories will flow to the IR fixed points related by the orientifold projection,  
but we know this should be the case thanks to the explicit construction of the fixed points. 
(It is possible that the large amount of symmetry helps to avoid possible corrections.)  
It strongly suggests that the orientifold equivalence between ABJM and ABJ holds even in this region. 
Indeed, at $k\ll N\ll k^5$ and $N\gg k^5$, 
we can use the IIA superstring and M-theory descriptions to show the equivalence, 
along the lines of  \cite{Hanada:2011yz} (Fig.~\ref{fig:equivalence_IIA},Fig.~\ref{fig:equivalence_M})\footnote{
Here we assume a stronger version of the Maldacena conjecture, which claims the gravity description is valid even outside the planar region 
as long as the stringy correction to the background metric is small, is correct.  
There are several observations supporting this assumption, including  
the Monte Carlo data from the D0-brane matrix quantum mechanics \cite{Hanada:2008gy,Hanada:2009ne} 
and exact calculation of BPS observables based on the localization method \cite{Pestun:2007rz,Kapustin:2009kz}. 
}. 
Then it is natural to expect the equivalence to hold in the intermediate region $N\sim k^5$ ($\lambda\sim N^{4/5}$). 
Actually, evidence for the equivalence can already be seen in the calculation of the free energy 
\cite{Drukker:2010nc,Fuji:2011km,Marino:2011eh,Hanada:2012si,Herzog:2010hf} based on the localization method 
\cite{Kapustin:2009kz},  it was found that there is no singularity around $N\sim k^5$ and 
the free energy of the $O(2N)_{2k}\times USp(2N)_{-k}$ theory is half of the free energy of the $U(2N)_{2k}\times U(2N)_{-2k}$ theory, as expected from a  ${\mathbb Z}_2$ projection\footnote{
Similar calculations \cite{Honda:2012ni} suggest that the Eguchi-Kawai equivalence for the supersymmetric Chern-Simons-matter theories 
\cite{Hanada:2009hd,Ishiki:2009vr} can be extended to the M-theory region. 
}. 
We further argue that the combination of the mirror symmetry and planar orientifold equivalence between mirrors provides us with other equivalences between more generic quiver theories.   

As a byproduct, we can reproduce a curious relation found in \cite{Aharony:2008gk};   
in the M-theory region, in addition to the orbifold equivalence explained here, 
there is an equivalence between two ABJM theories~\cite{Hanada:2011yz,Hanada:2011zx}
\begin{align} 
& U(2N)_{2k}\times U(2N)_{-2k}\to U(N)_{4k}\times U(N)_{-4k}, \label{EQU10}  
\end{align}
as mentioned above,  
and by combining \eqref{EQU9} and \eqref{EQU10} one obtains
\begin{align}
&U(N)_{4k}\times U(N)_{-4k}\to O(2N)_{\pm 2k}\times USp(2N)_{\mp k}. \label{EQU11}
\end{align} 
For $k=1$, this equivalence \eqref{EQU11} represents the quantum mechanical duality between the $U(N)_{4}\times U(N)_{-4}$ ABJM theory and the ABJ theory as pointed out in~\cite{Aharony:2008gk} and can exist in the presence of discrete holonomy of the 3-form potential. 
Although the equivalence \eqref{EQU9} holds both in IIA and M regions, \eqref{EQU10} holds only in the M-theory region and hence the equivalence \eqref{EQU11} as well. 

At first sight, the equivalence \eqref{EQU9} looks surprising from a string theory point of view; 
it seems as if  
a nice property of the classical type IIA string ('t Hooft limit) survives at quantum string level, 
after summing up the string loop corrections ($1/N$ corrections) to all orders. 
Probably, however, quantum string corrections do not play an important role; 
actually, 
in the explicit solution to the ABJM free energy \cite{Drukker:2010nc,Fuji:2011km,Marino:2011eh,Hanada:2012si,Herzog:2010hf}, 
which is obtained by using the localization method, 
the higher genus terms (higher orders in $g_{st}$) do not involve higher enough powers of $\lambda$ 
to compensate the suppression due to $g_{st}$, and hence only the planar diagrams survive even when $\lambda$ grows with $N$. Indeed the free energy takes the same form in IIA- and M-regions \footnote{
The leading part is $\frac{\sqrt{2}\pi}{3}\frac{N^{2}}{\sqrt{\lambda}}$ in the IIA limit. 
Although there is a correction of the form $\sum_{g=0}^\infty c_g (N^2/\lambda^2)^{1-g}$ \cite{Hanada:2012si,Drukker:2010nc,Marino:2011eh}, 
where $c_g$ are constants, 
that gives at most a constant contribution in the M-theory limit. Therefore the leading term remains the same, $\frac{\sqrt{2}\pi}{3}\frac{N^{2}}{\sqrt{\lambda}}
=\frac{\sqrt{2}\pi}{3}\sqrt{k}N^{3/2}$. 
This is consistent with a prediction from the gravity side \cite{Aharony:2008ug}. 
} 
. 
If this is really the case, the planar large-$N$ equivalence, and also other beautiful properties in the planar limit, 
can naturally be generalized, which would make studies of the classical M-theory within reach.\footnote{
This reminds us of the fact that the $1/N$ expansion makes sense in the strong coupling limit of the lattice gauge theory \cite{Wilson:1974sk}; 
note that this limit, in which the lattice coupling is sent to infinity for each fixed $N$, is similar to our limit.  
In early days of the study of large-$N$, 
based on the observation at strong coupling, quite a few people speculated that the planar calculation is valid even outside the 't Hooft limit. 
However at that time there was no explicitly calculable example. 
We thank H.~Kawai for enlightening comments on this point. 
} 
It is very interesting to study whether this property holds in other theories. 
Although direct test of the equivalence between supersymmetric Chern-Simons-matter theories would be difficult except for BPS sector 
where the localization method is applicable, in \S~\ref{sec:SYM} we argue that a similar equivalence can hold between certain 
Yang-Mills theories, for which full numerical simulation is applicable. 

The content of this paper is as follows: in \S~\ref{sec:projection_ABJM} we explain the orientifold projection from the perspective of the ABJM field theory. 
In \S~\ref{sec:brane_construction} and \S~\ref{sec:gravity} we give the brane constructions and   
gravity duals of the ABJM and ABJ theories, respectively, and show the orientifold equivalence. In particular, in \S~\ref{sec:fractional} we show the equivalence for gauge groups with different ranks. 
In \S~\ref{sec:mirror} we extend the equivalence to the mirror quiver theories. 
In \S~\ref{sec:SYM} we argue the same equivalence can hold in certain Yang-Mills theories and their orbifold/orientifold daughters. 
\section{Orientifold projection from  ABJM to ABJ}\label{sec:projection_ABJM}
We start by describing the orientifold projection of the ABJM theory with gauge group $U(2N)\times U(2N)$ to the ABJ theory with $O(2N)\times USp(2N)$ group. The field content of ABJM consists of two ${\cal N} = 2$ $U(2N)$ vector multiplets, an adjoint chiral multiplet for each gauge group and four chiral multiplets in the bifundamental representation, that we will denote as $A_a$, $B_a$, $a=1,2$. The action includes a Chern-Simons term for the gauge fields and its supersymmetric completion and after integrating out the adjoint chiral multiplet, a superpotential for the bifundamental multiplets \cite{Aharony:2008ug,Hosomichi:2008jb} 
\ba
W=\dfrac{2}{k}\mbox{tr}(A_{1}B_1A_2B_2-A_1B_2A_2B_1),
\ea
where we absorbed the $2\pi$ factors in $k$ compared with the normalization of~\cite{Aharony:2008ug}. Remember that this superpotential is obtained by using the similar method in~\cite{Klebanov:1998hh}  where the superpotential is obtained via the RG flow of a $d=4$ $\mathcal{N}=2$ gauge theory.

The orientifold projection acts differently on the two $U(2N)$ gauge groups, projecting one to an orthogonal $O(2N)$ group and the other to a unitary symplectic group $USp(2N)$. Denoting the gauge field of $O(2N)$ by $A_{\mu}$, the gauge field of $USp(2N)$ by $\tilde{A}_{\mu}$, and the scalar components of the bifundamental fields by $\Phi_{\alpha}=(\bar{B}_2,A_1,A_2,\bar{B}_1)$, the projected fields satisfy the relations 
\ba
&A_{\mu}=-A_{\mu}^T,\quad \tilde{A}_{\mu}=-J\tilde{A}_{\mu}^TJ^{-1},
&\bar{\Phi}_{\alpha}=(C_{\alpha}{}^{\beta}J\Phi^T_{\beta}), \label{ORI211}
\ea
where $J$ is the antisymmetric invariant tensor of $USp(2N)$ which satisfies $J^2=-1$. The antisymmetric tensor $C_{\alpha}{}^{\beta}$ is defined as $i\sigma_2\otimes 1$. The projection of the fermionic components is similar to that of scalars $\Phi_{\alpha}$. The condition on the scalars can also be expressed as
\ba
A_1=B_2^TJ,\quad A_2=-B_1^TJ.
\ea

The action for the fields in the projected theory is obtained directly by projecting the original ABJM action \cite{Aharony:2008ug}.
After the projection the superpotential becomes 
\ba
W=\dfrac{2}{k}\mbox{tr}(A_1JA_1^TA_2JA_2^T-A_1JA_2^TA_2JA_1^T).
\ea
The kinetic term for the gauge fields is a Chern-Simons term, originally with opposite levels $k$ and $-k$ for the two $U(2N)$ gauge groups. Using \eqref{ABC495}, the Chern-Simons action for the $O(2N)$ group becomes 
\ba
\notag &\dfrac{k}{2}\epsilon^{\mu\nu\rho}\mbox{tr}\Big(A_{\mu}\partial_{\nu}A_{\rho}+\dfrac{2}{3}A_{\mu}A_{\nu}A_{\rho}\Big) 
=\dfrac{k}{2}\epsilon^{\mu\nu\rho}\mbox{tr}\Big(A_{\mu}\partial_{\nu}A_{\rho}+\dfrac{1}{3}A_{\mu}[A_{\nu},A_{\rho}]\Big) \\
&=\dfrac{kC(G)}{2}\epsilon^{\mu\nu\rho}\Big(A_{\mu}^a\partial_{\nu}A_{\rho}^a+\dfrac{i}{3}f^{abc}A_{\mu}^aA_{\nu}^bA_{\rho}^c\Big). \label{CSA28}
\ea
And similarly for the $USp(2N)$ group. Our conventions regarding group theory factors $C(G)$ are explained in Appendix \ref{app}. After the orientifold projection, the normalization \eqref{NOR18} is such that the level of the $O(2N)$ group coincides with the normalization of $U(2N)$ for large values of $N$. However, for the $USp(2N)$ group the normalization \eqref{NOR19} implies that the level is halved. 
The covariant derivative of matter fields is also changed by the projection. Using \eqref{ORI211},  
\ba
D_{\mu}\bar{\Phi}_{\alpha}=JC_{\alpha}{}^{\beta}(D_{\mu}\Phi_{\beta})^T.
\ea
Comparing the projected expression with the ABJ $O(2N)\times USp(2N)$ action in \cite{Hosomichi:2008jb} one can check that both agree. Therefore, the projection we have described indeed corresponds to \eqref{EQU9}. Note that $USp(2N)\times O(2N)$ theory has ${\cal N}=5$ supersymmetry and an $SO(5)_R$ R-symmetry group that can be seen as a subgroup of the $SU(4)_R\times U(1)_b$ $R$-symmetry of the original ABJM theory. The orientifold projection removes the $U(1)_b$ baryonic symmetry.


The large-$N$ equivalence can be proven in the 't Hooft limit $N\to \infty$ and $\lambda=N/k$ fixed by using standard field theory methods 
\cite{Bershadsky:1998cb,Kovtun:2004bz}. 
\section{Type IIB brane construction}\label{sec:brane_construction}


\begin{figure}[htbp]
   \begin{center}
     \includegraphics[height=4cm]{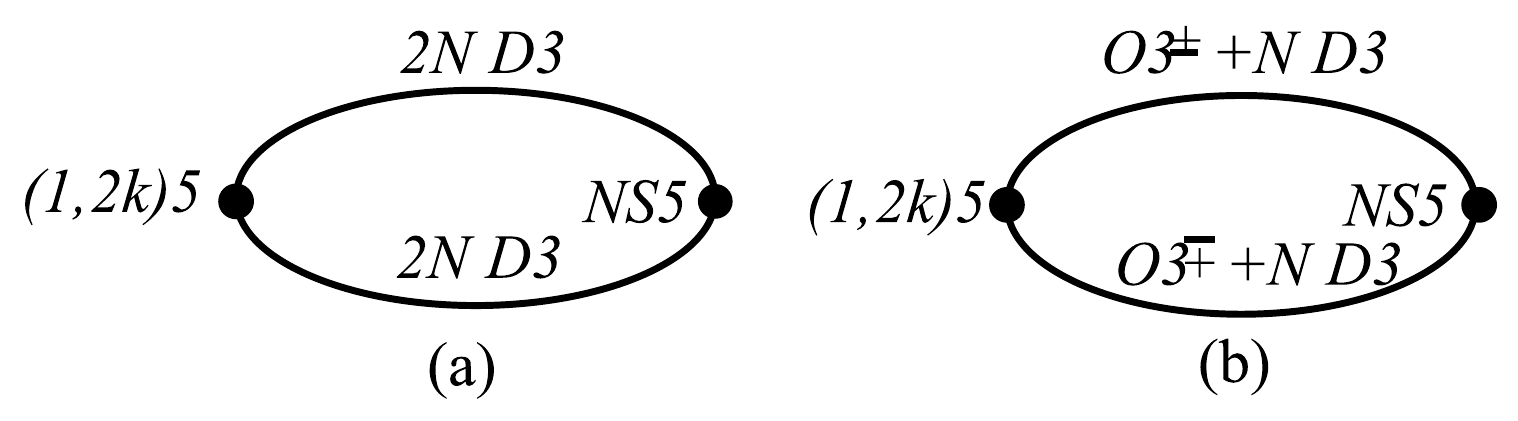}
   \end{center}
   \caption[orientifold]{(a) Type IIB elliptic brane configuration realizing the ABJM theory with gauge groups $U(2N)_{2k}\times U(2N)_{-2k}$. (b) Type IIB elliptic brane configuration realizing the ABJ theory with gauge groups $O(2N)_{\pm 2k}\times USp(2N)_{\mp k}$. 
 $N$ D3-branes are physical D3-branes and 5-branes are half 5-branes with their own images.}
\label{fig:orientifold}
\end{figure} 
The brane construction of the $U(2N)_{2k}\times U(2N)_{-2k}$ ABJM theory is obtained by including $2N$ D3-branes winding around a circle, intersecting with an NS5 and a $(1,2k)$5-brane at specific angles~\cite{Ohta,Bergman:1999na}\footnote{These 5-branes are linked with D3-branes in the context of the brane creation effect~\cite{Hanany:1996ie}.}. 
The  $O(2N)_{\pm 2k}\times USp(2N)_{\mp k}$ ABJ theory is constructed by adding to the ABJM construction $O3^{\pm}$ planes winding around a circle, in addition to the D3-branes and the two 5-branes (see Figure.~\ref{fig:orientifold}). The $2N$ D3-branes become $N$ physical branes plus their images, and 5-branes become half-branes with their own images also set on the top of the orientifold.
Depending on the orientifold plane, there is a different gauge group living on the D3-branes. For 
 $O3^{-}$, $O3^{+}$, $\tilde{O}3^{-}$, and $\tilde{O}3^+$ the gauge groups are $O(2N)$, $USp(2N)$, $O(2N+1)$, and $USp(2N)$, respectively~\cite{Hanany:2000fq,Feng:2000eq}. These four types of $O3$-planes are related by the $SL(2,\mathbb{Z})$ duality of Type IIB string theory, that also acts on $(p,q)$5-branes. 
 
In the ABJ construction there is a half NS5-brane and a half $(1,2k)$5-brane intersecting with the orientifold 3-plane. When the orientifolds cross a half NS5-brane they change their type, $O3^-$ changes to $O3^+$ and $O3^+$ changes to $O3^-$. Therefore, we have gauge groups $O(2N)\times USp(2N)$ on the D3 branes. Note that since the number of the half D5-branes is an even number $2k$, crossing them does not change the kind of $O3$ plane. One can change the relative rank of the groups by adding additional branes suspended between two 5-branes.

The number of D5 branes also determines the number of fundamental fields (flavors) that live on the D3 branes. Those are massive and can be integrated out, introducing a Chern-Simons term with a level proportional to the number of flavors. The $O(2N)$ Chern-Simons term has level $\pm 2k$, and the $USp(2N)$ Chern-Simons term has level $\mp k$. At low energies the Chern-Simons interaction dominates the dynamics and the theory flows to a fixed point.

From the perspective of the type IIB brane configurations, the equivalence between ABJM and ABJ theories can be seen as an ordinary orientifold equivalence between the four-dimensional theories living on the D3 branes, where the orientifold projection is due to the $O3$ planes. The equivalence is valid in the 't Hooft limit of the four-dimensional theory (not to be confused with the 't Hooft limit of the three-dimensional theory), when $N\to \infty$ and the four-dimensional 't Hooft coupling $g_{YM}^2N$ is fixed. 
At energies much below the size of the circle, the theory on the D3 branes becomes effectively three-dimensional.   
This three-dimensional theory flows to a fixed point at $E\ll g_{YM}^2k\sim k/N$. 
In the 't Hooft limit of the ABJM theory, where $\lambda=N/k$ is fixed, the fixed point can be in the planar region of the effective three-dimensional theory\footnote{
Note that the three-dimensional gauge coupling $g_{3d}^2$ has a dimension of mass. Therefore the planar scaling is realized when the dimensionless combination 
$g_{3d}^2N/E$, where $E$ is the energy scale under consideration, is of order one.  
}. Therefore it is natural to expect that IR fixed points of two theories are related by the planar orbifold equivalence of the UV theory. 
On the other hand, when $k$ is smaller than $O(N^1)$, the fixed point is outside the planar region of the UV theory,  
and so it is not guaranteed a priori that the two four-dimensional theories will flow to the IR fixed points related by the orientifold projection,  
but we know this should be the case thanks to the explicit construction of the fixed points. 
It is plausible that the large amount of symmetry helps to avoid possible corrections. 
Because both UV and IR theories are related by the orientifold projection, it is natural to expect the orientifold equivalence survives to IR, 
even when $\lambda=N/k$ is not of order one. Below we give argument supporting it. 

\section{Orientifold equivalence in the gravity dual}\label{sec:gravity}

From the brane configurations it is likely that the ABJM and ABJ theories are equivalent, 
since they can be seen as the low energy limit of two theories that are equivalent in the UV, 
and furthermore they are related by the same orientifold projection.   
We now provide stronger evidence, 
by showing how the gravity duals of the ABJM and ABJ fixed points \cite{Aharony:2008ug,Aharony:2008gk} 
are related in both eleven-dimensional supergravity and ten-dimensional type IIA supergravity \cite{Aharony:2008gk}. 
The AdS/CFT duality maps the equivalence in the gravity side to the gauge theory side (Fig.~\ref{fig:equivalence_IIA} and Fig.~\ref{fig:equivalence_M}). 

The space transverse to the $N$ M2-branes where the ABJM and ABJ theories live is $\mathbb{C}^4/\mathbb {Z}_{2k}$ and $\mathbb{C}^4/\hat{\mathbb{D}}_k$, respectively, where $\hat{\mathbb{D}}_k$ is a diehdral group with $4k$ elements. The dihedral group can be decomposed in a $\mathbb{Z}_{2k}$ action, that also appears in the ABJM case, and an additional $\mathbb{Z}_2$ action, that we can identify with the orientifold projection. When $N$ is large, the M2 branes backreact on the geometries and in the near horizon limit there is a dual description of the M2 branes as M-theory on the orbifold geometries $AdS_4\times S^7/{\mathbb Z}_{2k}$ and  
 $AdS_4\times S^7/\hat{\mathbb{D}}_k$, respectively. The M-theory description is valid at $N\gg k^5$, where the size of the M-theory circle is larger than the eleven dimensional Planck scale.   
At  $k\ll N\ll k^5$, the systems are well described by the type IIA supergravity on the orbifold geometries  $AdS_4\times \mathbb{CP}^3$ and $AdS_4\times \mathbb{CP}^3/\mathbb{Z}_2$, respectively.

Let us parametrize the space transverse to the M2 branes by the complex coordinates $z_i$ $(i=1,2,3,4)$. The $\mathbb{Z}_{2k}$ action of the orbifold is
\ba
z_i\to e^{i\frac{\pi}{k}}z_i. \label{ORB517}
\ea
To describe the additional $\mathbb{Z}_2$ action in $\hat{\mathbb{D}}_k$, we should write the $\mathbb{C}^4/{\mathbb{Z}}_{2k}$ space as a product of two Taub-NUT geometries. These are  hyper-K{\"a}hler manifolds and the center of each Taub-NUT geometry is locally a flat $\mathbb{C}^2$. We consider the following unit sphere in the $\mathbb{R}^3$ space that is the base of the Taub-NUT geometry:
\ba
&f:\mathbb{C}^2\to S^2, \\
&f(z_1,z_2)=(2\Re (z_1z_2^*),2\Im (z_1z_2^*),|z_1|^2-|z_2|^2),
\ea
and we can define a similar unit sphere for the other Taub-NUT factor. Recall that orientifolds in  string theory reverse the sign of the coordinates in $\mathbb{R}^3$ when described in terms of the type IIB theory. Thus, the $\mathbb{Z}_2$ action operates as the antipodal map on the $S^2$ inside the $\mathbb{C}^2$ at the center of each Taub-NUT and can be lifted to the action on $z_1,z_2$ and $z_3,z_4$. The action is
\ba
z_1\to iz_2^*,\quad z_2\to -iz_1^*,\quad z_3\to iz_4^*,\quad z_4\to -iz_3^*.\quad
\ea

To connect with the geometry that we use in the AdS/CFT duality, we write $\mathbb{C}^4/\hat{\mathbb{D}}_k$ as the cone over $S^7/\hat{\mathbb{D}}_k$. Here, $S^7/\hat{\mathbb{D}}_k$ is embedded in $z_i$ satsifying $\sum_{i=1}^4|z_i|^2=1$ as follows:\footnote{We follow the notation in \cite{NTP}.}
\begin{align}
z_1&=\cos\xi \, \cos\f{\theta_1}{2} \, e^{i\f{\chi_1+\vp_1}{2}} ~, \qquad
z_2=\cos\xi \,  \sin\f{\theta_1}{2} \, e^{i\f{\chi_1-\vp_1}{2}} ~, \nonumber \\
z_3&=\sin\xi \,  \cos\f{\theta_2}{2} \, e^{i\f{\chi_2+\vp_2}{2}} ~, \qquad
z_4=\sin\xi \, \sin\f{\theta_2}{2} \, e^{i\f{\chi_2-\vp_2}{2}} ~,
\label{angles}
\end{align}
where the ranges of the angular variables are $0\leq \xi
<\f{\pi}{2}$, $0\leq \chi_i <4\pi$, $0\leq \vp_i < 2\pi$ and $0\leq \theta_i<\pi$. 
The $\mathbb{Z}_{2k}$ orbifold action is taken along the
$y$-direction as $y\sim y+\f{\pi}{k}$, where the new coordinate $y$
is defined by
\begin{align}
\chi_1=2y+\psi ~, \qquad \chi_2=2y-\psi ~.
\end{align}
In addition, the $\mathbb{Z}_2$ action is operated on the angular variables as follows:
\ba
\theta_i \to \pi -\theta_i,\quad \varphi_i \to \varphi_i +\pi,\quad \chi_i \to -\chi_i. \label{Z205}
\ea
When the backreaction of the M2-branes is considered, the gravitatinal solution has $F_4$ flux and the geometry is changed to $AdS_4\times S^7/\hat{\mathbb{D}}_k$, where the compact part of the geometry should be identified with the original base of the cone.

In the absence of an orbifold $\mathbb{Z}_{2k}$, $y$ is replaced by a circle $y'$ and the gravity side is $AdS_4\times S^7$:
\ba
&ds^2_{11D}=\frac{R^2}{4}(ds^2_{AdS_4}+4ds^2_{S^7}),\quad
  ds^2_{S^7}=(dy'+A)^2+ds^2_{\mathbb{CP}^3}, \\
&N'=\frac{1}{(2\pi \ell_p)^6}\int_{S^7} \ast F_4,\ F_4=\frac{3}{8} R^3 vol_{AdS_4},  \\
&6R^6 vol(S^7)=2\pi^4R^6=(2\pi \ell_p)^6 N', \label{VOL617}
\ea
where using the coordinate \eqref{angles}, the gauge potential and the metric of $\mathbb{CP}^3$ is given by
\begin{align}
A=\f{1}{2}(\cos^2\xi-\sin^2\xi)d\psi +\f{1}{2}\cos^2\xi
\cos\theta_1 d\vp_1+ \f{1}{2}\sin^2\xi \cos\theta_2 d\vp_2 ~.\end{align}
\begin{align}
&ds^2_{\mathbb{CP}^3}=d\xi^2+\cos\xi^2\sin^2\xi\left(d\psi+\f{\cos\theta_1}{2}d\vp_1-\f{\cos\theta_2}{2}d\vp_2\right)^2 \no 
& +\f{1}{4}\cos^2\xi\left(d\theta_1^2+\sin^2\theta_1
d\vp_1^2\right)+\f{1}{4}\sin^2\xi(d\theta_2^2+\sin^2\theta_2
d\vp_2^2).
\end{align}
Here $N'=2Nk$. 
When the orbifold is introduced, the $\mathbb{Z}_{2k}$ quotient is performed over \eqref{VOL617} and the $\mathbb{Z}_2$ action \eqref{Z205} operates on the angular variables. That is, rewriting $y'\to y'/(2k)$ with $y'\sim y'+\pi$, the metric can be rewritten as
\ba
ds^2_{S^7/\mathbb{Z}_k}=\dfrac{1}{(4k)^2}(dy' +2kA)^2+ds^2_{\mathbb{CP}^3}. \label{FIB518}
\ea
From the volume formula \eqref{VOL617}, replacing $vol(S^7)$ with $vol(S^7/\hat{\mathbb{D}}_k)$, it can be shown that $R/l_p=(2^7\pi^2 kN)^{1/6}$. Here, remember that there is a tadpole cancellation between each $O3^{\pm}$-branes and so we do not need to take into account the $O3^{\pm}$-charges. 
In order for the classical M-theory description to be valid, size of the orbifolded M-theory circle must be larger than the eleven-dimensional Planck length, 
$(R/l_p)/2k\sim (kN)^{1/6}/k\gg 1$, and hence $N\gg k^5$ is required. 

Now let us consider the type IIA reduction. The radius of the $\mathbb{CP}^3$ metric in \eqref{FIB518} is large if $kN\gg 1$. However, the radius of $\varphi$ is of the order of $R/k\propto (Nk)^{1/6}/k$. So, the weakly coupled Type IIA theory description requires $k^5\gg N$. Using the ansatz of the dimensional reduction,
\ba
&ds^2_{11D}=G^{11}_{MN}(x^{\mu})dx^{M}dx^{N} \nonumber \\
&=e^{-\frac{2}{3}\phi}G^{10}_{\mu\nu}dx^{\mu}dx^{\nu}+e^{\frac{4}{3}\phi}(dy'+2kA)^2, \label{KK357}
\ea
 the IIA string frame metric gives
the $AdS_4\times \mathbb{CP}^3$ IIA background and the background flux:
\ba
&ds^2_{st}=L^2(ds^2_{AdS_4}+4ds^2_{\mathbb{CP}^3}), \nonumber \\
&e^{2\phi}= R^3 /(2k)^3,\ L^2= R^3/(8k)=\pi\sqrt{\dfrac{2N}{k}}, \nonumber \\
&F_2=\f{2k}{L^2}\omega
~,\ \ \   \ti{F}_4(\equiv F_4-C_1\we H_3)=-\f{3}{8}R^3 \ep_{AdS_4}
~,\ \ \ H_3=0 ~. \label{CPT518}
\ea
 where $\phi$ is the dilaton field and $\omega$ is the K{\"a}hler form of $\mathbb{CP}^3$. There is also an additional $\mathbb{Z}_2$ orientifold action relative to the original ABJM dual geometry.
 Note that the $F_4$ flux becomes $N=(\int_{\mathbb{CP}^3/\mathbb{Z}_2} *F_4/(2\pi)^5)$, consistent with the brane configuration where we normalize the flux to be an integer using the volumes of $\mathbb{CP}^1$ and $\mathbb{CP}^3$\footnote{ Also note that the volume of the unit $S^7$ is
$\mbox{Vol}(S^7)=\f{\pi^4}{3}$.} \be \mbox{Vol}(\mathbb{CP}^1) =4\pi
L^2,\quad \mbox{Vol}(\mathbb{CP}^3)=\f{32}{3}\pi^3
L^6. \ee
 (in this paper we always work with the string frame metric
setting $\al=1$). The $RR$ 2-form $F^{(2)}=2kdA$ in the Type IIA theory (we sometimes call this a D6-brane flux) is explicitly given as follows
\begin{eqnarray} F^{(2)}&=& 2k\Bigl(-\cos\xi\sin\xi d\xi \we
(2d\psi+\cos\theta_1d\vp_1-\cos\theta_2 d\vp_2)\no &&
-\f{1}{2}\cos^2\xi\sin\theta_1 d\theta_1\we d\vp_1
-\f{1}{2}\sin^2\xi\sin\theta_2 d\theta_2 \we d\vp_2\Bigr). \label{fluxsix}
\end{eqnarray}
We can show that the physical D6-brane flux becomes $k$ consistent with the brane configuration.

Then, the curvature radius $(=2^{5/2}\pi \sqrt{N/k})$  should be large so that the supergravity description is valid. Note that curvature radius of the $AdS_4\times \mathbb{CP}^3$ in type IIA theory becomes the same form $R^2_{string}=2^{5/2}\pi\sqrt{N'/k'}$ if we substitute $N=2N'$ and $k=2k'$.  
 Here, 't Hooft coupling of the ABJ theory is defined by $\lambda =N/k$. As a result, the type IIA description is valid in the regime  
\ba
k\ll N\ll k^5.
\ea
The orientifold action $\mathbb{Z}_2$ maps $\omega\to -\omega$ on $\mathbb{CP}^3$ and the orientation of $\mathbb{CP}^3$ is reversed. The orientifold also flips the sign of the RR 1-form $C_1$ and the NSNS 2-form $B$, while RR 3-form is invariant under this action. 

From this analysis it is clear that the dual geometry to the ABJ fixed point is simply related to the ABJM one by the additional $\mathbb{Z}_2$ action in the $\hat{\mathbb{D}}_k$ orbifold, that in the type IIA limit becomes the orientifold action. The equivalence can be formulated for any observable that is invariant under the $\mathbb{Z}_2$ projection, its value computed using classical supergravity or other classical objects like the DBI action in string theory should  give the same results in both geometries. A similar statement can be made between ABJM theories with different orbifold actions \cite{Hanada:2011yz,Hanada:2011zx}.
\subsection{Orientifold theories with fractional branes}\label{sec:fractional} 
We can extend the arguments of the previous section to cases where the rank of the gauge groups in the ABJM and ABJ theory are not equal, the difference coming from the introduction of fractional branes  \cite{Aharony:2008gk}. 

\begin{figure}[htbp]
   \begin{center}
     \includegraphics[height=3cm]{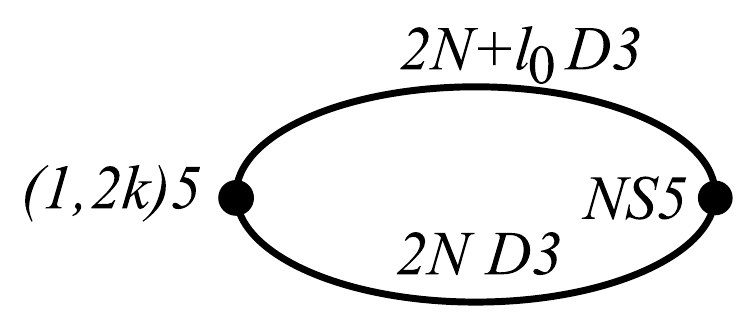}
   \end{center} 
    \caption[fractional2]{The Type IIB elliptic brane configuration of the ABJ theory with the gauge group $U(2N+l_0)_{2k}\times U(2N)_{-2k}$. }
\label{fig:fractionalD3}
\end{figure} 

The type IIB brane construction of the ABJM theory includes $2N$ D3-branes winding around a circle, intersecting an NS5 and a $(1,2k)$5-brane at specific angles~\cite{Ohta,Bergman:1999na}. A $U(2N+l_0)_{2k}\times U(2N)_{-2k}$ theory is obtained if $l_0$ D3-branes are suspended between the NS5-brane and the $(1,2k)$5 on one side of the circle. See Figure. \ref{fig:fractionalD3}. In this construction, the classical moduli space is identical to the moduli space of the $U(2N)_{2k}\times U(2N)_{-2k}$ corresponding to the motion of the $2N$ free D3-branes, there is no moduli space associated with $l_0$ locked D3-branes. 

Performing a $T$-duality transformation on the type IIB brane configuration with $l_0$ fractional branes and lifting  to M-theory, one obtains a configuration with $N$ M2-branes on a cone $\mathbb{C}^4/\mathbb{Z}_{2k}$  
 plus $l_0$ fractional M2-branes at the orbifold singularity. Here, the fractional M2-branes correspond to the discrete torsion~\cite{WB} realized by a discrete holonomy of the 3-form potential $\exp (i\int_{S^3/\mathbb{Z}_{2k}}C_3)\in \mathbb{Z}_{2k}$.  This discrete holonomy implies that $2k$ wrapped fractional M2-branes are equivalent to none.  After including the backreaction of $2N$ branes and taking the near horizon limit, we obtain the metric $AdS_4\times S^7/\mathbb{Z}_{2k}$ with discrete torsion:
\ba
&ds^2_{11D}=\frac{R^2}{4}(ds^2_{AdS_4}+4ds^2_{S^7/\mathbb{Z}_{2k}}),\quad R/l_p=(2^7\pi^2kN)^{1/6}.
 \label{VOLapp}
\ea

\begin{figure}[htbp]
   \begin{center}
     \includegraphics[height=4cm]{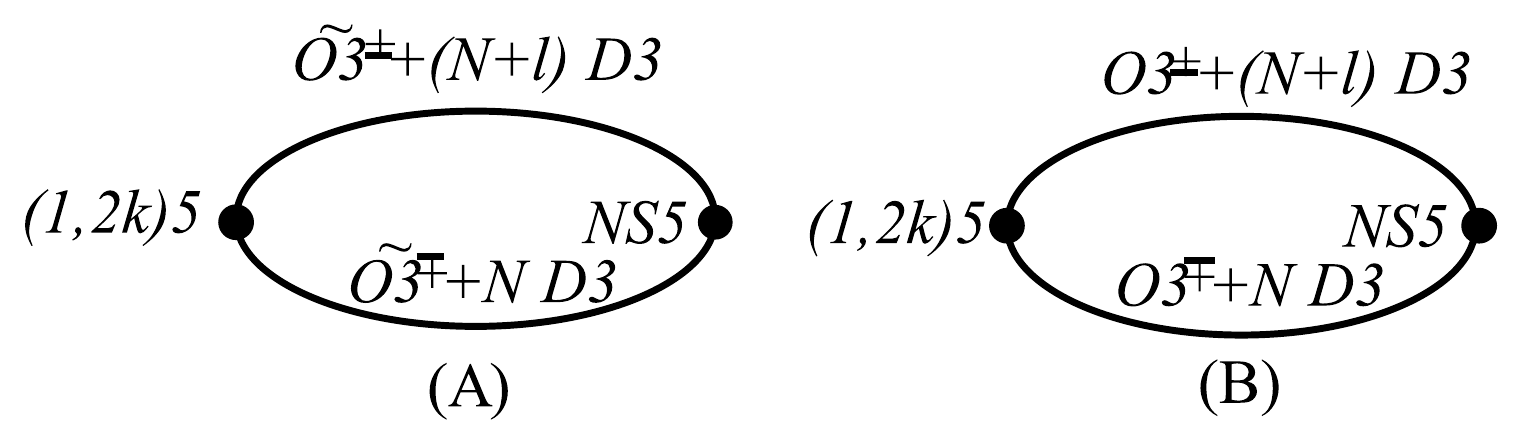}
   \end{center} 
    \caption[fractional]{The Type IIB elliptic brane configuration of the ABJ theory with orientifold planes. There are 4 classes of orientifold theories. We set $l_0=2l$ compared with Figure. 5.}
\label{fig:fractional}
\end{figure} 

The Kaluza-Klein reduction to type IIA is performed in the same way as in the $l_0=0$ case. Recall that the type IIA theory description is valid for $N^{1/5}\ll k \ll N$. The $AdS_4\times \mathbb{CP}^3$ background metric is
\ba
&ds^2_{st}=L^2(ds^2_{AdS_4}+4ds^2_{\mathbb{CP}^3}),\quad L^2= R^3/(8k)=\pi\sqrt{\dfrac{2N}{k}}. 
\ea
The dual of a $U_{2k}(2N+l_0)\times U_{-2k}(2N)$ theory should have a background 2-form flux $B_2$ associated to the discrete torsion. The M-theory 3-form reduces to the 2-form $B_2$, which gives a non-trivial holonomy on $\mathbb{CP}^1$ in $\mathbb{CP}^3$, $b_2=\int_{\mathbb{CP}^1}B_2/(2\pi)^2=\f{l_0}{2k}-\f{1}{2}$, where a shift of $1/2$ is included in fluxes of $B_2$ as found in~\cite{Aharony:2009fc}. Then, $B_2$ is quantized in units of $1/2k$.


There are also $2k$ units of flux $F_2$ on $\mathbb{CP}^1$. This flux changes the Bianchi identity of $\tilde{F_4}$ to $d\tilde{F}_4=-F_2\wedge H_3=-d(F_2\wedge B_2)$. Such identity implies that the conserved flux is not $\int F_4$ but $\int \tilde{F}_4/(2\pi)^3 = 2kb_2$.

We now introduce the orientifold 3-planes in the original type IIB setup. Remember that there are 4 orientifold planes $O3^-,O3^+,\tilde{O3}^{-},\tilde{O3}^{+}$. Considering 
the $SL(2,\mathbb{Z})$ duality of type IIB string theory, we can find four classes of orientifold theories in Figure. \ref{fig:fractional}. Here, we introduced $l_0=2l$ fractional branes which are consistent with the $\mathbb{Z}_2$ orientifold. The gauge symmetry of these theories are given by
\ba
&(1): \ O(2N+2l+1)_{2k}\times USp(2N)_{-k} \quad (0\le l< k),\nonumber \\
&(2): \ USp(2N+2l)_{k}\times O(2N+1)_{-2k} \quad (0\le l< k), \nonumber \\
&(3): \ O(2N+2l)_{2k}\times USp(2N)_{-k} \quad (0\le l< k+1), \nonumber \\
&(4): \ USp(2N+2l)_{k}\times O(2N)_{-2k} \quad (0\le l< k-1).
\ea
As seen in the case without fractional branes, the matter content is determined by using the orientifold projection of 
$U(2N)_{2k}\times U(2(N+l))_{-2k}$ bifundamental fields 
\ba
\bar{\Phi}_{\alpha}=(C_{\alpha}{}^{\beta}J\Phi^T_{\beta}),
\ea
where $J$ and the antisymmetric tensor $C_{\alpha}{}^{\beta}$ were given in \eqref{ORI211}. The superpotential of the $U(2N)_{2k}\times U(2(N+l))_{-2k}$ theory is projected in the same way as in the case without fractional branes. 
The restriction for the number of fractional branes is can be checked from the s-rule and the brane creation~\cite{Bergman:1999na,Hanany:1996ie,Feng:2000eq}:
In total, there are $4k$ different theories.

In the gravity dual for the orientifold theory the  $S^7/\mathbb{Z}_{2k}$ factor of the geometry is replaced  by
$S^7/\hat{\mathbb{D}}_k$. There is a 3-cycle in $S^7/\hat{\mathbb{D}}_{k}$ with discrete holonomy given by
$\exp(i\int_{3-cycle}C_3)=\mathbb{Z}_{4k}$
where $l=0,...,4k-1$. Thus, there are $4k$
different theories classified by $H_3(S^7/\hat{\mathbb{D}}_k,\mathbb{Z}) = \mathbb{Z}_{4k}$ consistent with the Type IIB brane configurations. Considering the backreaction of M2-branes and taking the near horizon limit, we obtain the gravity theory on $AdS_4\times \hat{\mathbb{D}}_k$ with discrete holonomy, the metric is
\ba
&ds^2_{11D}=\frac{R^2}{4}(ds^2_{AdS_4}+4ds^2_{S^7/\hat{\mathbb{D}}_{k}}),\quad R/l_p=(2^7\pi^2kN)^{1/6}.
 \label{VOLapp2}
\ea
Remember that the AdS radius does not change in the presence of the holonomy, this implies that there is a large-$N$ equivalence between theories with different $l$. In the M-theory regime this is in agreement with our expectations, since $l\ll N$. 


We now perform the Kaluza-Klein reduction to type IIA theory. It is similar to the case without orientifold planes except for the $\mathbb{Z}_2$ orbifolding. The metric of the ten-dimensional Type IIA theory is given by
\ba
&ds^2_{st}=L^2(ds^2_{AdS_4}+4ds^2_{\mathbb{CP}^3/\mathbb{Z}_2}),\quad L^2= R^3/(8k)=\pi\sqrt{\dfrac{2N}{k}}. 
\ea
 Remember that the orientifold action does not flip the sign of $C_3$ but flips the sign of $C_1$ and $B_2$, while  in addition makes a tensor transformation of these fields. To include the fractional D2-brane flux that makes the rank of two gauge groups different, we define the discrete holonomy $b\equiv\int_{\mathbb{CP}^1/\mathbb{Z}_2}B_2/(2\pi)^2=\f{l}{4k}$. Here, a shift of $1/2$ needs to be included in the fluxes of $B_2$ as seen in the original type IIB setup. The discrete torsion is then reduced to $\int_{\mathbb{CP}^2}\tilde{F}_4=4kb$ and there are $4k$ possible discrete holonomy in terms of the NSNS 2-form. Since the metric is unaffected by the discrete torsion, there is also a $\mathbb{Z}_2$ equivalence in the type IIA regime between the $4k$ classes of ABJM theory with gauge group $U(2N)_{2k}\times U(2(N+l))_{-2k}$ and the corresponding ABJ theory obtained through the orientifold projection.

\section{Mirror brane configurations}\label{sec:mirror}
\begin{figure}[htbp]
   \begin{center}
     \includegraphics[height=6cm]{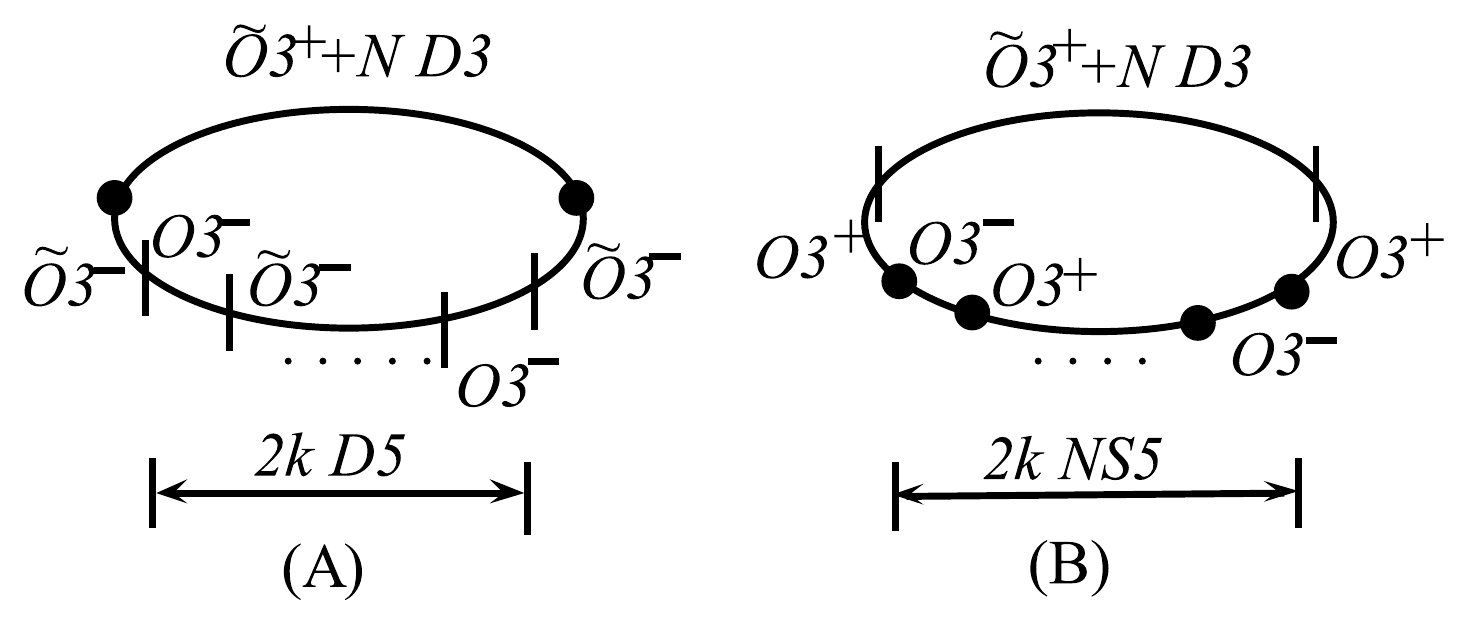}
   \end{center}
   \caption[mirror]{(A) shows the elliptic D3-brane configuration before the S-duality. $N$ physical D3-branes are placed on any intervals of the circle. A large dot describes a half NS5-brane and a vertical line describes a half D5-brane.  (B) shows the mirror configuration of (A). $N$ physical D3-branes are placed on any intervals of the circle.}
\label{fig:mirror}
\end{figure}

In this section we combine the orbifold equivalence with mirror symmetry, in this way one can derive other nontrivial equivalences. Note that mirror symmetry is a consequence of S-duality in the type IIB D-brane construction that describes the UV theory, so it takes a theory with coupling $g_{YM}$ to a theory with coupling $1/g_{YM}$. This also gives an indication that orbifold equivalences extend beyond the planar limit, since in the regime $g_{YM}^2\sim N$ the UV theory can be mapped to a theory in the 't Hooft limit where the equivalence can be proved by the usual means. The main caveats concern the low energy limit of the theories, either when $k\ll N$ is small and the IR fixed points are at very low energies $E/\lambda_{3d}\sim 1/N$  or when $k\sim N$ and there is a large number of massless states in the mirror theory.

\begin{figure}[htbp]
   \begin{center}
     \includegraphics[height=4cm]{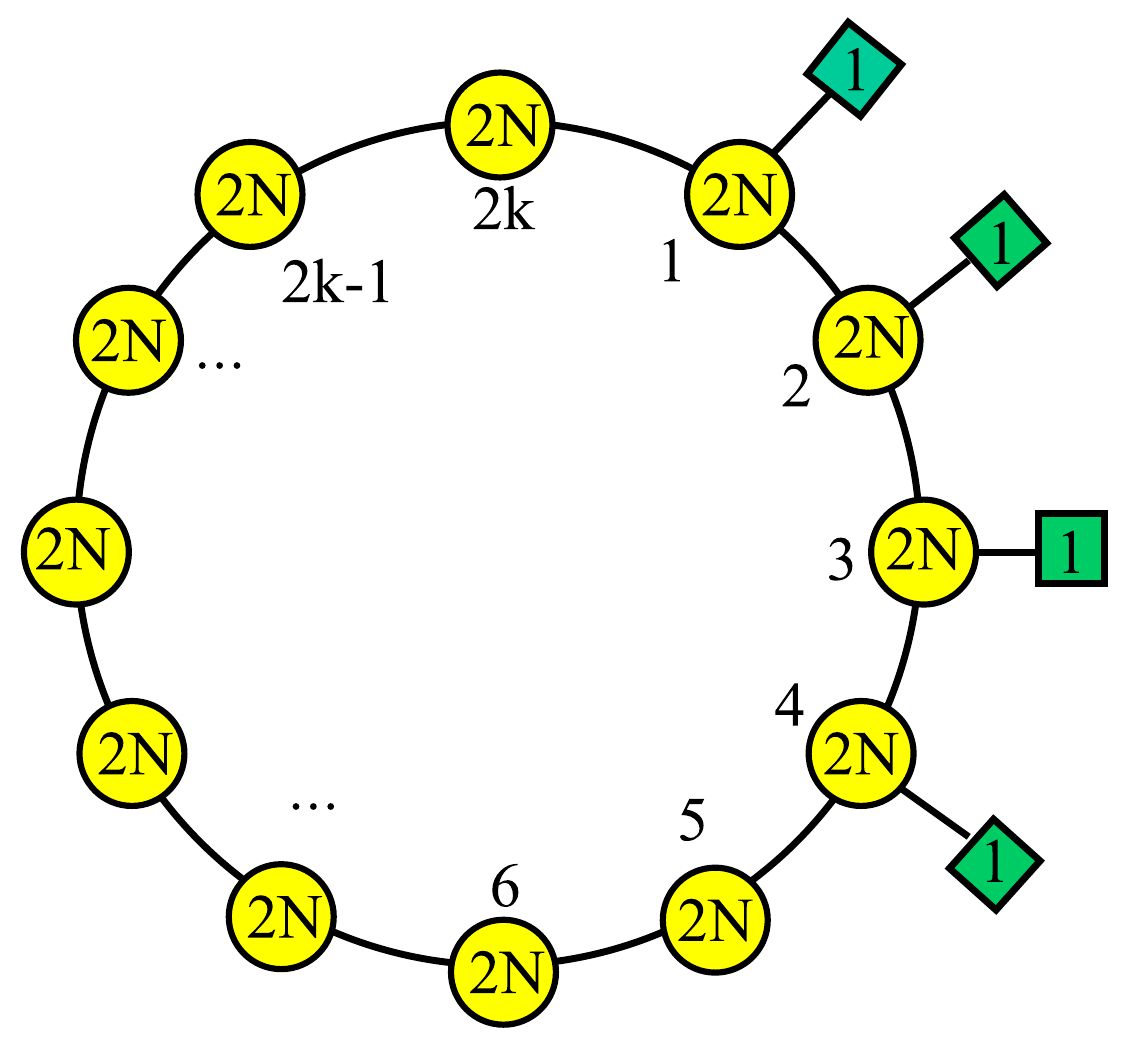}
   \end{center}
   \caption[quiver]{Quiver diagram of the mirror side of ABJM and ABJ. 
   The mirror of ABJM is a $(U(2N)\times U(2N))^k$ quiver gauge theory with four fundamental hyper multiplets. 
   The nodes (yellow disks) represents the $U(2N)$ groups and links connecting the nodes represent the bifundamental matters as usual.  
  Each green box represents a global flavor group, and links connecting the nodes and the boxes are the fundamental hyper multiplets. 
   For ABJ, nodes labeled by the odd number and the even number describe the gauge group $O(2N)$, $USp(2N)$, respectively. 
   Here, the third node describes $O(2N+1)$ instead of $O(2N)$. }
\label{fig:quiver}
\end{figure}

In terms of string theory, the mirror symmetry is the S-duality in type IIB brane configurations. 
The mirror dual of the $U(2N)_{2k}\times U(2N)_{-2k}$ ABJM theory has been considered in \cite{Jensen:2009xh} 
and used in \cite{Hanada:2011zx} to study the M-theory region. 
At low energy the mirror is a $(U(2N)\times U(2N))^k$ quiver gauge theory with four fundamental hyper multiplets (Figure.~\ref{fig:quiver}). 

In order to obtain the mirror to the $O(2N)_{2k}\times USp(2N)_{-k}$ ABJ theory, we start from the brane construction \cite{Aharony:2008gk}, 
that we described at the beginning of section~\ref{sec:brane_construction}, 
where D3 branes and O3$^{\pm}$ planes are along 0126 directions, NS5-branes along 012345 directions and D5-branes along 012349 directions 
(Figure.~\ref{fig:mirror}(A)). Note that the O3-plane also changes its type crossing either D5-brane or NS5-brane for above brane configurations. This is because forgetting the NS5-brane once and considering our D3/D5 system, the discrete charge of D5-brane changes when O3-plane crosses the D5-brane. The same analysis can be applied if we  consider the D3/NS5 system forgetting the D5-brane.

We perform an S-duality and obtain a configuration with $2k$ half NS5-branes along 012578 directions and two half D5-branes along 012789 directions 
(Figure.~\ref{fig:mirror}(B)). We consider the cross configuration where each half D5-brane is on top of a single half NS5-brane~\cite{Brunner:1998jr,Park:1999eb}.  This configuration preserves $d=3$ $\mathcal{N}=2$ supersymmetry and for a single half NS5-brane, there are two copies of a fundamental half hypermultiplet for the two gauge groups via the flavor doubling. There also appears two global flavor symmetry associated with these hypermultiplets. 
Remember that in general, two 5-branes are linked if there is only a transverse direction to both the D5-brane and the NS5-brane.
 Since the half D5-brane is not linked with the half NS5-brane in our case, there is not the brane creation by definition.

Note however that $O3^+(\tilde{O}3^+)$ turns to $O3^-(\tilde{O}3^-)$ when it crosses the (half-)NS5-brane and vice versa. In addition, $O3^-(O3^+)$ turns to $\tilde{O}3^+(\tilde{O}3^-)$ when it crosses a half D5-brane on top of a half NS5-brane and vice versa, so we obtain a chain of gauge groups $(O(2N)\times USp(2N))^{k-1}\times O(2N+1)\times USp(2N)$, where a gauge group of $N$ D3-branes on $\tilde{O}3^-$ becomes $O(2N+1)$ instead of $O(2N)$. See Figure. \ref{fig:quiver}.
 It can be shown that this change is consistent with the S-duality of O3-planes which transforms  $O3^-,O3^+,\tilde{O}3^{-},\tilde{O}3^+$ into $O3^-,\tilde{O}3^-,O3^+,\tilde{O}3^+$, respectively.  

We can obtain the same quiver by doing the orientifold projection of $(U(2N)\times U(2N))^k$, which is the mirror to $U(2N)_{2k}\times U(2N)_{-2k}$. 
Before the orientifold projection $\Omega_1 =\Omega (-1)^{F_L}\gamma_{345789}$, the matter content consists of $2k$ chiral multiplets in the adjoint $Y_i$, $2k$  hypermultiplets transforming in the bifundamental representation of $(i,i+1)$ groups $(A_{i,i+1},B_{i+1,i})$, two chiral multiplets transforming in the fundamental representation under the first and second gauge group ${D}_1,U_2$, and two chiral multiplets transforming in the anti-fundamental $\tilde{U}_1,\tilde{D}_2$.  There are also four (anti-)fundamental chiral multiplets from the $j$-th and $(j+1)$-th gauge group $D_j,U_{j+1},\tilde{U}_j,\tilde{D}_{j+1}$, where subscripts of these fields describe the corresponding gauge group factor. We choose $D_3,U_{4}$,$\tilde{U}_3,\tilde{D}_{4}$ for $j=3$ to realize the quiver diagram \ref{fig:quiver}.\footnote{The case for $D_2,U_{3},\tilde{U}_2,\tilde{D}_{3}$  $(j=2)$ is also interesting since it corresponds to the configuration where the D5-branes are aligned adjacently.}

The orientifold projection $\Omega_1$, imposes the following conditions on the fields
\ba
\begin{cases} B_{i+1,i}=JA_{i,i+1}^T,\quad \text{for $i$ odd}, \\
B_{i+1,i}=-A_{i,i+1}^TJ,\quad \text{for $i$ even}, \end{cases} \label{ORI236}
\ea
and
\ba
\tilde{D}_{i}=U_{i}^T,\quad {D}_{i+1}=-J\tilde{U}_{i+1},\quad \tilde{D}_{i+1}=U_{i+1}^TJ,\quad D_i=\tilde{U}_i^T,\quad (i=1,3) \label{ORI237}
\ea  
where we take the convention that if $i$ is odd, the $i$-th gauge group is $O(2N)$ (the third gauge group is $O(2N+1)$) and if $i$ is even, it is $USp(2N)$.
We summarize the field content in the quiver diagram in Figure. \ref{fig:quiver}. There is the $\mathcal{N}=2$ superpotential with 2 cubic interaction terms before the orientifold action as follows:
\ba
&W_0=\sum_{i=1,3}\Big[\tilde{U}_iA_{i,i+1}U_{i+1}-\tilde{D}_{i+1}B_{i+1,i}D_i\Big]. 
\ea
 The $\mathcal{N}=2$ superpotential of the orientifold theory is then projected into 1 cubic interaction term as follows:
\ba
&W=2\sum_{i=1,3}\tilde{U}_iA_{i,i+1}U_{i+1}.
\ea

The quiver theory in the mirror side can also be obtained from an orbifold projection. We start with $U(2kN)\times U(2kN)$ theory with 4 hypermultiplets where $k$ is an odd integer. The matter content  consists of $2$ chiral multiplets in the adjoint $Y_i$, $2$  hypermultiplets transforming in the bifundamental representation of $(i,i+1)$ groups $(A_{i,i+1},B_{i+1,i})$, four chiral multiplets transforming in the fundamental representation under the first and second gauge group ${D}_{(a)1},U_{(a)2}$, and four chiral multiplets transforming in the anti-fundamental $\tilde{U}_{(a)1},\tilde{D}_{(a)2}$ where $a=1,2$. The $\mathcal{N}=2$ superpotential becomes
\ba
S=\sum_{a=1}^2\Big[\tilde{U}_{(a)1}A_{1,2}U_{(a)2}-\tilde{D}_{(a)2}B_{2,1}D_{(a)1}\Big].
\ea
Recall that in terms of the fields in the ABJ theory, the bifundamental matter is represented by $(A_{1,2},B_{2,1})=(A_1,B_2)$ and $(A_{2,1},B_{1,2})=(B_1,A_2)$. There is an enhanced $(U(1)\times U(1))^2$ global symmetry.

 The $\mathbb{Z}_k$ orbifold projection is obtained from the element of each gauge group $U(2kN)$ and spans a $\mathbb{Z}_k$ subgroup as follows:
\ba
\gamma = \text{diag}(1_{\mathbf{2N}},\ \omega 1_{\mathbf{2N}},\ \omega^2 1_{\mathbf{2N}},\ ...\ ,\omega^{k-1} 1_{\mathbf{2N}}),
\ea
where $1_{2\mathbb{N}}$ is the $2N\times 2N$ identity matrix and we have defined the phase $\omega =e^{2\pi i/k}$. $k$ should be odd since for even $k$, the quiver diagram is separated into two parts as also observed in the case of orbifolds of the ABJM theory~\cite{Terashima:2008ba}.

We consider the orientifold action $\mathbb{Z}_N+\Omega_1\mathbb{Z}_N$ where the orientifold $\Omega_1=\Omega (-1)^{F_L}\gamma_{345789}$ is defined in \eqref{ORI236} and \eqref{ORI237}~\cite{Park:1999eb}. 
The quiver gauge theory is obtained from the $U(2kN)\times U(2kN)$ theory by keeping the components that are invariant under the following projection:
\ba
&V_i\to \gamma V_i\gamma^{-1},\quad Y_i\to \gamma Y_i\gamma^{-1}, \\
&A_{i,i+1}\to \omega \gamma A_{i,i+1}\gamma^{-1},\quad B_{i+1,i}\to \omega^{-1}\gamma B_{i+1,i}\gamma^{-1}, \\
&\tilde{U}_{(a)1}\to \omega^{2a-2}\tilde{U}_{(a)1}\gamma^{-1},\quad U_{(a)2}\to \omega^{1-2a}\gamma U_{(a)2}, \quad (a=1,2), \\
&\tilde{D}_{(a)2}\to \omega^{2a-1} \tilde{D}_{(a)2}\gamma^{-1},\quad D_{(a)1}\to \omega^{2-2a}\gamma D_{(a)1}, \quad (a=1,2),
\ea
If we apply now the orientifold action $\Omega_1$,  we obtain the quiver gauge theory of the gauge group $(O(2N)\times USp(2N))^{k-1}\times O(2N+1)\times USp(2N)$. The other quiver gauge theories with flavors coupling with different nodes are obtained by using the different orbifold condition and by coupling the flavors with $A_{2,1},B_{1,2}$. We can summarize the orbifold projections in the mirror theories as:
\ba
&U(2kN)\times U(2kN)+ 4\ \text{hypermultiplets} \label{GRA544}  \\
& \downarrow \ \text{Orbifold projection} \nonumber \\
&(U(2N)\times U(2N))^k+ 4\ \text{hypermultiplets}   \\
& \downarrow \ \text{Orientifold projection} \nonumber \\
&(O(2N)\times USp(2N))^{k-1}\times O(2N+1)\times USp(2N)+ 4\ \text{half hypermultiplets} \nonumber \\
& \updownarrow \ \text{Mirror} \nonumber \\
&O(2N)\times USp(2N)+ 4k\ \text{half hypermultiplets}, \nonumber\\
& \uparrow \ \text{Orientifold} \nonumber \\
&U(2N)\times U(2N)+ 4k\ \text{half hypermultiplets}. \nonumber
\ea
The flow to the IR fixed point described by ABJM theory is not directly given by the brane configurations we have considered. The field content is the same, but one should add a mass deformation for the fields in the fundamental representation in the original theory $O(2N)\times USp(2N)$. As seen in ~\cite{Hanada:2011zx}, in the mirror theory $(O(2N)\times USp(2N))^{k-1}\times O(2N+1)\times USp(2N)$, the mass deformation maps to some non-local deformation such as the monopole operator.

\section{Equivalences between Yang-Mills theories}\label{sec:SYM}
So far we have studied examples involving supersymmetric Chern-Simons theories in the large-$N$ limit. 
Do similar equivalences beyond the 't Hooft limit exist in Yang-Mills theories? 

As a concrete example, let us consider the three-dimensional ${\cal N}=8$ $U(kN)$ super Yang-Mills theory (SYM), the low energy description of $kN$ D2 branes at the origin of the moduli space. This is the UV description of the $U(kN)_1\times U(kN)_{-1}$ ABJM theory. 
In this theory the coupling constant $g_{YM}^2$, and hence the 't Hooft coupling $\lambda_{YM}=g_{YM}^2N$,  
has the dimension of mass, and it sets an energy scale of the theory. The planar description is valid when 
$\lambda_{YM}/E=O(1)$, where $E$ is the energy scale under consideration. At very long distance, $E\ll g_{YM}^2$, 
the ABJM theory should give a good description. Obviously, this limit is different from the 't Hooft limit. 

Let us consider ${\mathbb Z}_k$ orbifold projections of this theory, which utilizes the $SO(7)$ R-symmetry. 
For simplicity we consider the ones which are obtained from the familiar projections in four-dimensional ${\cal N}=4$ super Yang-Mills theory considered in \cite{Kachru:1998ys,Douglas:1996sw}, that is, we consider a ${\mathbb Z}_k$ transformation of the form 
\begin{eqnarray}
Z_1\to e^{2\pi i n_1/k} Z_1, 
\qquad
Z_2\to e^{2\pi i n_2/k} Z_2, 
\qquad
Z_3\to e^{2\pi i n_3/k} Z_3, 
\end{eqnarray}
where $Z_{1,2,3}$ are complex scalars which describe six of seven transverse coordinates to the D2 branes. 
According to a stronger version of the gauge/gravity conjecture, the IIA supergravity description is expected to be valid at $1\ll \lambda/E\ll N^{4/5}$ \cite{Itzhaki:1998dd},   
although the relationship between the $1/N$ expansion and $g_{string}$ expansion is not clear unless $\lambda/E$ is of order $N^0$. 
By assuming it, one can easily see the orbifold equivalence in this region. 
At further lower energy 
it is natural to expect that the IR fixed point of this orbifold daughter is described by M-theory 
on an orbifold of $AdS_4\times S^7$.\footnote{ 
It is not clear whether the IR fixed point is described by an orbifold projection of the $U(kN)_1\times U(kN)_{-1}$ ABJM theory, 
which has been considered in \cite{Benna:2008zy,Terashima:2008ba};  
actually a usual logic \cite{Douglas:1996sw}, combined with the quantization of the level, 
seems to require the parent's level is multiple of $k$, in order for the daughter to admit a usual interpretation of the moduli space 
as a multiple M2-branes on top of the orbifold singularity \cite{Terashima:2008ba}. 
Hence the projection becomes like 
\begin{eqnarray}
U(kN)_{kl}\times U(kN)_{-kl}
\longrightarrow
\left(U(N)_l\times U(N)_{-l}\right)^k,  
\end{eqnarray}
where $l$ is integer, which prevents us from starting with $U(kN)_1\times U(kN)_{-1}$. 
Still it is plausible that the IR fixed point admits a gravitational description, and that is the only assumption needed for our discussion.  
We thank F.~Yagi for a very useful discussion on this issue. 
}  
Similarly to the example discussed above, this projection should give the equivalence in the M-theory region. 
Then we have the ${\mathbb Z}_k$ orbifold equivalence of three-dimensional SYM in the UV region (which is just the usual orbifold equivalence) 
and deep IR (which lies outside the 't Hooft limit). It strongly suggests that this equivalence holds at any energy scale. 

The same can hold in other gauge theories. 
For four-dimensional ${\cal N}=4$ SYM and its orbifold/orientifold daughters, the IIB supergravity description is expected to be valid at $1\ll \lambda\ll N$ \cite{Itzhaki:1998dd},  
and hence planar equivalence should extend to that region.  
It is possible that the equivalence can be generalized to $\lambda\gg N$ by using S-duality. 
The same argument applies to theories in $0+1$ and $1+1$ dimensions as well. 
Simple tests in the BPS sector should be possible 
by using the localization method (see e.g. \cite{Pestun:2007rz,Okuda:2010ke}). 
A possible equivalence can also be tested by Monte Carlo simulation.\footnote{ 
In $(0+1)$-dimension, because simulation cost is not very expensive, detailed simulation can be performed (see \cite{Hanada:2007ti} and following works). In $(1+1)$-dimension, lattice formulations keeping a few exact supersymmetries \cite{Cohen:2003xe} 
turned out to be free from the parameter fine tuning even nonperturbatively \cite{Kanamori:2008bk} 
and hence test of the equivalence is within reach. (Actually two-dimensional lattice has already been used to learn about very interesting physics, 
namely the string theory in D1-brane background \cite{Catterall:2010fx}.) 
$(1+2)$- and $(1+3)$-dimensional maximally supersymmetric theories can be studied by utilizing a fine-tuning free formulation utilizing 
fuzzy spheres \cite{Maldacena:2002rb} \cite{Hanada:2010kt}, although it is not easy to go to larger matrix size 
with current numerical resources.  
Lattice formulation may also work, as suggested in \cite{Catterall:2011pd}, because a number of fine-tunings can be small. 
(Also the Eguchi-Kawai approach \cite{Ishii:2008ib} might work if it is valid outside the 't Hooft limit as suggested in \cite{Honda:2012ni}.)
} 
In fact there is an observation which supports the equivalence:  in the one-dimensional theory (D0-brane quantum mechanics), 
a class of two-point functions seems to agree with the predictions from IIA supergravity, even in the M-theory region   
\cite{Hanada:2009ne}, and hence the orbifold equivalence will hold as well.  

Although we have focused on supersymmetric theories, it will also be interesting to consider nonsupersymmetric theories.  
Probably the simplest setup for a check of orbifold equivalences is the two-dimensional pure Yang-Mills with $SU(2N)$, $SO(2N)$ and $USp(2N)$ gauge groups, 
which can be studied both analytically \cite{Migdal:1975zg,Gross:1993hu}\footnote{
Calculation of the free energy of the dimensionally reduced model \cite{Ishiki:2008vf} suggests the validity of the Eguchi-Kawai reduction outside the planar limit in this case. 
} and numerically. We believe it would be very interesting to pursue this direction further. 
\section{Discussion and outlook}
There are many posible future directions one can pursue. First of all, it is important to understand under which conditions the equivalence can be extended beyond the 't Hooft limit. 
In the case of the ABJM theory we could show the equivalence at any strong coupling because the orientifold in the string theory region 
naturally lifts to an orbifold in the M-theory region. In many other theories, however, the orbifold/orientifold projections in the gravity duals do not have such 
natural lifts, and then the equivalences could be justified only in the string theory region\footnote{
Note however that the string region already contains a part of the very strongly coupled limit. 
}.  
However, the existence of such lifts could just be a sufficient condition, 
and the equivalence might still extend outside the string region. One natural possibility is that the equivalence holds unless a phase transition 
separates the 't Hooft limit and the very strongly coupled region. 
It would be interesting to construct such examples with and without a phase transition, in order to test this scenario.     
It is also interesting to see whether the equivalence outside the planar limit 
can hold in nonsupersymmetric theories. For that purpose, two-dimensional pure Yang-Mills should be a good laboratory. 
Also it is important to test the simplest scenario: planar dominance outside the planar large-$N$ limit. 
For that purpose, numerical checks of the factorization in pure Yang-Mills would be the easiest approach.  
If this simple scenario is correct, other nice properties of the planar limit, for example the integrability, might be generalized to the new large-$N$ limit. 
It would be fascinating because it might enable us to study the M-theoretic aspects of the AdS/CFT correspondence.  
The equivalence between supersymmetric Chern-Simons theories itself is also important to pursue further, 
because it might be useful to gain insights into condensed matter systems; 
by turning the table around, 
it might provide an `experimental test' of the orbifold equivalence, 
if theories related by the orbifold equivalence can be realized in a laboratory. 
Supersymmetric Chern-Simons theories which are constructed by the low energy limit of the type IIB brane configurations 
and have the gravity dual~\cite{Imamura:2008nn,Imamura:2008dt,Jafferis:2008qz} are probably the easiest examples to study. 
In the paper~\cite{Terashima:2008ba}, $\mathbb{Z}_n$ orbifolds of the ABJM theory are studied and the M2-brane theories on a $\mathbb{C}^4/\mathbb{Z}_{kn}\times \mathbb{Z}_{n}$ singurality are proposed. The $AdS_4\times S^7/(\mathbb{Z}_{kn}\times \mathbb{Z}_{n})$  gravity dual of this orbifolded theory is also studied in~\cite{Imamura:2008ji}.  
So, it implies that we can propose the orbifold equivalence between the $U(nN)_{nk}\times U(nN)_{-nk}$ ABJM theory and the quiver CSM type with the product gauge group $(U(N)_{k}\times U(N)_{-k})^n$. It is also interesting to study the orbifold equivalence between $d=3$ $\mathcal{N}=3$ quiver Chern-Simons-matter theories including flavors which have the corresponding gravity dual~\cite{Gaiotto:2009tk,Hohenegger:2009as,Hikida:2009tp,Fujita:2009xz}\footnote{
See also \cite{Armoni:2008kr}, although an existence of an M-theory description is not clear.  
}.  

\section*{Acknowledgements}
The authors would like to thank A.~Karch for stimulating discussions, comments, and collaboration at early stage of this work. 
The authors would also like to thank T.~Azeyanagi, M.~Buchoff, A.~Cherman, S.~Cremonesi, Y.~Hikida, M.~Honda, K.~Hosomichi, H.~Kawai, K.~Murakami, H.~Shimada, S.~Shimasaki, F.~Sugino, S.~Terashima,  
M.~Unsal and F.~Yagi for useful comments and discussions.  
M.~H. is grateful to the hospitality of the Kavli Institute for Theoretical Physics in UCSB, 
during the KITP program ``Novel Numerical Methods for Strongly Coupled Quantum Field Theory and Quantum Gravity". 
M.~F. is supported from Postdoctoral Fellowship for Research Abroad by Japan Society for the Promotion of Science. This research was  supported in part by the National Science Foundation under Grant No. PHY11-25915.
This work was supported in part by the U.S. Department of Energy under grant DE-FG02-96ER40956. 
C.~H. was supported in part by the Israel Science Foundation (grant number 1468/06).

\appendix
\section{Group theory conventions}\label{app}

We use a basis of generators normalized as\footnote{
We follow the notation of Peskin's textbook. 
} 
 \ba
\mbox{tr}(t^a_rt^b_r)=C(r)\delta^{ab},
\ea
where $C(r)$ describes a constant for each representation $r$. Above the equation and the commutation relation $[t_r^a,t_r^b]=if^{abc}t_r^c$ gives the following representation of the structure constants: 
\ba
f^{abc}=-\dfrac{i}{C(r)}\mbox{tr}\{[t^a_r,t_r^b]t^c_r\}. \label{ABC495}
\ea
This formula shows that $f^{abc}$ is totally antisymmetric.

The product of $t^{a}_r$ summed over the index $a$ is proportional to the unit matrix
\ba
t^a_rt^a_r=C_2(r)\cdot 1, \label{CA496}
\ea
where 1 is the $d(r)\times d(r)$ unit matrix and $C_2(r)$ describes the quadratic Casimir for each representation. If we contract \eqref{ABC495} with $\delta_{ab}$ and calculate the left-hand side using \eqref{CA496}, we find 
\ba
d(r)C_2(r)=d(G)C(r). \label{REL497}
\ea
A convention of $C(r)$ is $C(N)=\eta/2$ for the generators of $SU(N)$. 
For the fundamental representation $N$ and $\bar{N}$, $C_2(N)$ is derived from  
 \eqref{REL497}
 \ba
 C_2(N)=\dfrac{N^2-1}{N}C(N).
 \ea

To compute the Casimir for the adjoint representation, the product of the $N$ and $\bar{N}$ representations is used.
For $SU(N)$, 
\ba
C_2(G)=C(G)=2NC(N).
\ea
Symmetric and antisymmetric tensors form irreducible representations of $SU(N)$. 
The direct sum of these representations is the product representation $N\times N$.
The relation of such traces between antisymmetric, symmetric representation is
\ba
\mbox{tr}_A(t^at^b)=(N-2)C(N)\delta^{ab},\quad \mbox{tr}_S(t^at^b)=(N+2)C(N)\delta^{ab}. \label{NOR18}
\ea

For $SO(N)$, normalizing differently in terms of $C(N)=\eta$~\cite{vanRitbergen:1998pn}, 
\ba
C_2(N)=(N-1)C(N)/2 \quad C(G)=C_2(G)=(N-2)C(N).
\ea

For $USp(2N)$, using the same normalization of $C(2N)=\eta/2$ as $SU(N)$
\ba
C_2(N)=\dfrac{1}{2}C(N)(2N+1),\quad C(G)=C_2(G)=2C(N)(N+1). \label{NOR19}
\ea
In the main section, we should set $\eta =1/(2N)$ in \eqref{CSA28} to be consistent with the normalization in 
 the brane configurations.

\end{document}